\documentclass[%
notitlepage,%
nofootinbib,%
amsmath,%
amssymb,%
superscriptaddress,%
]{revtex4-1}

\usepackage[utf8]{inputenc}
\usepackage[english]{babel}
\usepackage{graphicx}
\usepackage{caption}
\usepackage{subcaption}

\begin{document}
\title{Spatial structure of the modified Coulomb potential\\
  in a superstrong magnetic field}

\author{S.I. Glazyrin}
\email{glazyrin@itep.ru}
\affiliation{Dukhov Research Institute of Automatics (VNIIA),
  ul. Sushchevskaya 22, Moscow, 127055 Russia}
\affiliation{Alikhanov Institute for Theoretical and Experimental
  Physics, National Research Centre Kurchatov Institute, ul. Bol’shaya
  Cheremushkinskaya 25, Moscow, 117218 Russia}
\affiliation{National Research Nuclear University MEPhI, Kashirskoe
  sh. 31, Moscow, 115409 Russia}
\affiliation{Novosibirsk State University, Novosibirsk, 630090,
  Russia}

\author{S.I. Godunov}
\email{sgodunov@itep.ru}
\affiliation{Alikhanov Institute for Theoretical and Experimental
  Physics, National Research Centre Kurchatov Institute, ul. Bol’shaya
  Cheremushkinskaya 25, Moscow, 117218 Russia}
\affiliation{Novosibirsk State University, Novosibirsk, 630090,
  Russia}

\begin{abstract}
  The modification of the Coulomb potential due to the enhancement of loop
  corrections in a superstrong magnetic field is studied
  numerically. We calculate the modified potential with high precision
  and obtain the pattern of equipotential lines.  The results confirm
  the general features known from previous studies, however we
  emphasize some differences in potential structure that can be
  important for problems with spatially distributed charges.
\end{abstract}

\maketitle

\section{Introduction}
\label{sec:intro}

The Coulomb potential is responsible for the appearance of the bound
states of oppositely charged particles. The behaviour $U\propto 1/r$
originate from a single photon exchange. With the account for
radiative corrections the potential will be different from the Coulomb
one but the deviations are usually small. In the presence of the
external fields these corrections can be enhanced.

It is well known that vacuum polarization in one loop is enhanced by
an external magnetic field (proportional to magnetic field $B$). It
was discovered by Shabad and Usov \cite{ShU-1,ShU-2} that this
enhancement leads to the significant modification (screening) of the
Coulomb potential for $B\gtrsim m^{2}/e^{3}$ where $m$ and $e$ are
electron mass and charge\footnote{The Gauss system of units is used,
  $e^{2}=\alpha=1/137.035\dots$, so $m^{2}/e^{3}=B_{0}/\alpha$ where
  $B_{0}\equiv m^{2}/e\approx 4.4\cdot 10^{13}~\mbox{G}$.}. They
calculated the potential in cylindrical geometry ($\rho$, $\phi$, $z$)
numerically for $z=0$ with arbitrary $\rho$ and for $\rho=0$ with
arbitrary $z$ (the pointlike charge is located at
$\left(\rho,~z\right)=(0,~0)$ and magnetic field is directed along $z$
axis) and found asymptotics analytically.

Interpolation (but still quite accurate) formula for the polarization
operator in a superstrong magnetic field was suggested by Vysotsky
\cite{Vysotsky-2010}. With the help of this formula the potential for
$z=0$ (arbitrary $\rho$) and for $\rho=0$ (arbitrary $z$) was
calculated analytically \cite{MV,GMV}. It was found
\cite{ShU-1,ShU-2,GMV} that for $z\gg 1/m$ the equipotential lines are
ellipses, though the potential in the mid-range distances,
$r=\sqrt{\rho^{2}+z^{2}}\lesssim 1/m$, was still unknown. In Fig. 1 in
\cite{GMV} the equipotential lines were approximated by ellipses
everywhere.

The aim of the present paper is to find out what happens with the
potential at mid-range distances. To do that we evaluate the modified
potential numerically. The equipotential lines at distances
$r\sim 1/m$ turn out to be ``eye-shaped'' (see
Fig. \ref{fig:lines_1e4}, \ref{fig:lines_1e5}) rather than
elliptic. It means that the potential diminishes with $\rho$ faster
than it was expected earlier. Such result should be important for
problems with the distributed charges, e.g. for the calculation of
energy levels in the field of a nucleus with the finite radius.

In Section \ref{sec:analytical} we reproduce basic formulae from
\cite{ShU-1,ShU-2,Vysotsky-2010,MV,GMV}. In Section
\ref{sec:numerical} we present the results of numerical
calculations. We conclude in Section \ref{sec:conclusions}.

\section{Modified potential}
\label{sec:analytical}

Let us briefly describe current analytical results on the modified
potential.  In the presence of an external superstrong magnetic field
($B\gg B_{0}\equiv m^{2}/e\approx 4.4\cdot 10^{13}~\mbox{G}$) the
contribution of vacuum polarization at one loop level to the
polarization operator becomes greatly enhanced \cite{LS}:
\begin{equation}
  \label{eq:VP}
  \Pi^{(2)}\left(k_{\perp},k_{\parallel}\right)=-\frac{2e^{3}B}{\pi}\exp\left(-\frac{k_{\perp}^{2}}{2eB}\right)T\left(t\right),
\end{equation}
where
$k=\left(0,\vec k\right)=\left(0,\vec{k}_{\perp},k_{\parallel}\right)$
is the momentum of the external photon,
$t\equiv{k_{\parallel}^{2}}/{4m^{2}}$ and
\begin{equation}
  \label{eq:T}
  T(t)=1-\frac{1}{\sqrt{t(1+t)}}\log\left(\sqrt{1+t}+\sqrt{t}\right).
\end{equation}
Here we took into account only Lowest Landau Level (LLL) contribution
into polarization operator which should dominate. Contributions from
higher loops are also omitted in what follows. With these
approximations we follow
\cite{ShU-1,ShU-2,Vysotsky-2010,MV,GMV,LS,Godunov-2loop} where the
arguments in favour of this approximation can be found.

Since the polarization operator enters the photon propagator, it leads
to the modification of the pointlike charge potential (we consider
elementary charge $e$):
\begin{equation}
  \label{eq:potential}
  \Phi\left(\rho,z\right)= 4\pi
  e\int\frac{d^{2}k_{\perp}dk_{\parallel}}{\left(2\pi\right)^{3}}
  \frac{e^{-i\vec{k}_{\perp}\vec\rho}e^{-ik_{\parallel}z}}{k_{\parallel}^{2}+k_{\perp}^{2}-\Pi^{(2)}\left(k_{\perp},k_{\parallel}\right)}.
\end{equation}

After integration over the angle in the plane transverse to the
magnetic field the following result was obtained \cite{ShU-1,ShU-2}
(up to units and notations):
\begin{equation}
  \label{eq:full}
  \Phi(\rho,z)=\frac{e}{\pi}\int\limits_{-\infty}^{\infty}dk_{\parallel}e^{-ik_{\parallel}z}
  \int\limits_{0}^{\infty}dk_{\perp}
  \frac{k_{\perp}J_{0}(k_{\perp}\rho)}{k_{\perp}^{2}+k_{\parallel}^{2}+\frac{2e^{3}B}{\pi}e^{-k_{\perp}^{2}/2eB}T(k_{\parallel}^{2}/4m)}.
\end{equation}

In \cite{Vysotsky-2010} the interpolation formula for $T(t)$ was
introduced:
\begin{equation}
  \label{eq:interpolation}
  T(t)\approx \frac{2t}{2t+3}.
\end{equation}
It is rather simple and allows to perform analytical calculations, but
at the same time it provides a very good accuracy (see \cite{MV} for
details).

Using formula (\ref{eq:interpolation}) the analytical expressions for
$\Phi\left(0,z\right)$ and $\Phi\left(\rho,0\right)$ were found
\cite{MV,GMV} (see also asymptotics in \cite{ShU-1,ShU-2}):
\begin{equation}
  \label{eq:Phi_z}
  \Phi\left(0,z\right)=\frac{e}{|z|}\left(1-e^{-|z|\sqrt{6m^{2}}}+e^{-|z|\sqrt{(2/\pi)e^{3}B+6m^{2}}}\right),
\end{equation}
and for $B\gg 3\pi m^{2}/e^{3}$:
\begin{equation}
  \label{eq:Phi_rho}
  \Phi\left(\rho,0\right)=\left\{
      \begin{array}{ll}
        \frac{e}{\rho}\cdot\exp\left(-\rho\sqrt{(2/\pi)e^{3}B}\right),& \rho<l_{0}, \\
        \frac{e}{\rho}\cdot\sqrt{\frac{3\pi m^{2}}{e^{3}B}},& \rho>l_{0},
      \end{array}
    \right.
\end{equation}
where
$l_{0}\equiv\sqrt{\frac{\pi}{2e^{3}B}}\ln\sqrt{\frac{e^{3}B}{3\pi
    m^{2}}}$.

The potential at large distances, $z\gg 1/m$, was found in
\cite{ShU-1,ShU-2,GMV}:
\begin{equation}
  \label{eq:Phi_far}
  \left.\Phi\left(\rho,z\right)\right|_{z\gg 1/m}=
  \frac{e}{\sqrt{z^{2}+\rho^{2}\left(1+\frac{e^{3}B}{3\pi m^{2}}\right)}},
\end{equation}
which means that potential lines at large distances are ellipses.

The equipotential lines were shown in Fig. 1 in \cite{GMV} where they
were found with the help of (\ref{eq:Phi_z}), (\ref{eq:Phi_rho}), and
(\ref{eq:Phi_far}). The equipotential lines were supposed to be
ellipses everywhere. The aim of this paper is to lift this assumption
and to calculate initial integral~(\ref{eq:full}) in all space
numerically. This is done in Section \ref{sec:numerical}.

\section{Numerical results}
\label{sec:numerical}

We want to calculate the potential numerically with the maximum
achievable precision\footnote{Let us note that there
  are contributions to the potential other than from vacuum
  polarization at one loop with electrons at LLL. So generally
  speaking, it is pointless to infinitely hunt for the precision in
  one particular contribution. But it is important for us to check
  that we can achieve high precision so in future we can take into
  account other effects as well.}. In order to do that one should
calculate (\ref{eq:full}). But the integral has a singularity at
$k_{\perp}=0,~k_{\parallel}=0$ and therefore it is not really suitable
for numerical evaluation.

This singularity has exact physical meaning and it is due to
singularity in unmodified Coulomb potential, so we subtract $e/r$ from
the integral to regularize it.  The Coulomb potential can be
represented in the form same to (\ref{eq:full}) but without
$\Pi^{(2)}\left(k_{\perp},k_{\parallel}\right)$ in the
denominator. Therefore the difference is\footnote{The same trick was
  used in \cite{ShU-1}.}
\begin{equation}
  \label{eq:diff}
  \Delta\Phi\left(\rho,z\right)\equiv\frac{e}{\sqrt{\rho^{2}+z^{2}}}-\Phi(\rho,z)=
  \frac{e}{\pi}\int\limits_{-\infty}^{\infty}dk_{\parallel}e^{-ik_{\parallel}z}
  \int\limits_{0}^{\infty}dk_{\perp}
  k_{\perp}J_{0}(k_{\perp}\rho)
  \frac{\frac{2e^{3}B}{\pi}e^{-k_{\perp}^{2}/2eB}T(k_{\parallel}^{2}/4m)}
  {\left(k_{\perp}^{2}+k_{\parallel}^{2}\right)\left(k_{\perp}^{2}+k_{\parallel}^{2}+\frac{2e^{3}B}{\pi}e^{-k_{\perp}^{2}/2eB}T(k_{\parallel}^{2}/4m)\right)}.
\end{equation}
This expression is finite for any $\rho$ and $z$ though for numerical
integration some additional regularizations are required. The
expression for T(t), Eq.~(\ref{eq:T}), should be expanded into series
for small $t$, and the numerator and denominator under the integral
should be divided by $k_{\parallel}^{2}$ for small
$k_{\parallel}$. Numerically we integrate in two steps: first
integration over $k_{\perp}$ is taken with the help of GNU Scientific
Library \cite{GSL}, second integration over $k_\parallel$, the Fourier
transform, is provided by the FFTW package \cite{FFTW}. We carefully
estimated the errors of the integrations: the final result absolute
error for $B=10^{4}B_{0}$ is less than $10^{-7}(m\cdot e)$ for any
$\rho$ and $z$. For $B=10^{5}B_{0}$ the error is less than
$10^{-6}(m\cdot e)$. Our numerical results are in a good agreement
with numerical results for $\Phi\left(0,z\right)$ and
$\Phi\left(\rho,0\right)$ obtained in \cite{ShU-1,ShU-2}.

\begin{figure}[ht!]
  \centering
  \begin{subfigure}[b]{0.49\textwidth}
    \includegraphics[width=\textwidth]{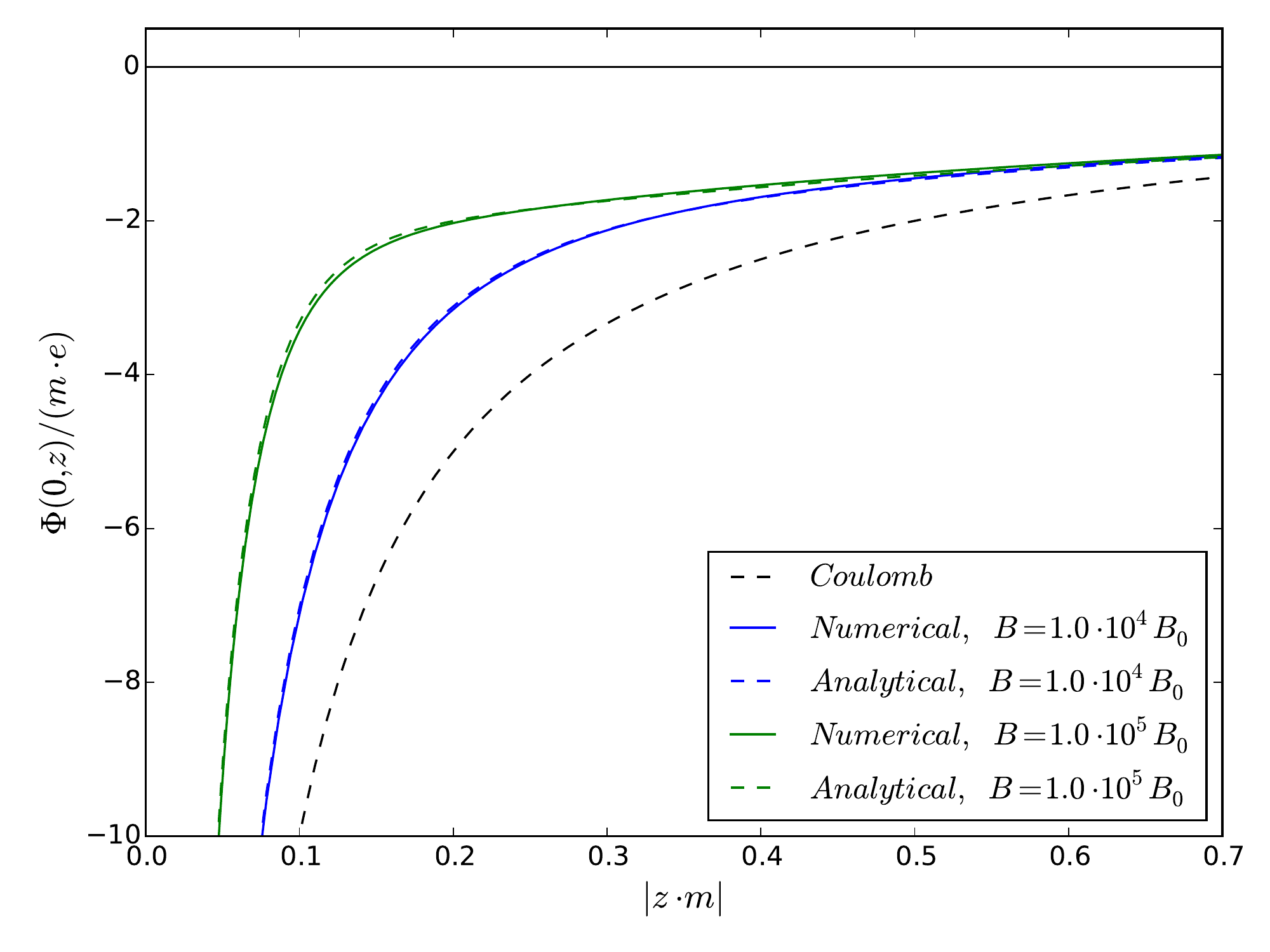}
    \caption{$\Phi\left(0,z\right)$}
    \label{fig:phi_z}
  \end{subfigure}
  \begin{subfigure}[b]{0.49\textwidth}
    \includegraphics[width=\textwidth]{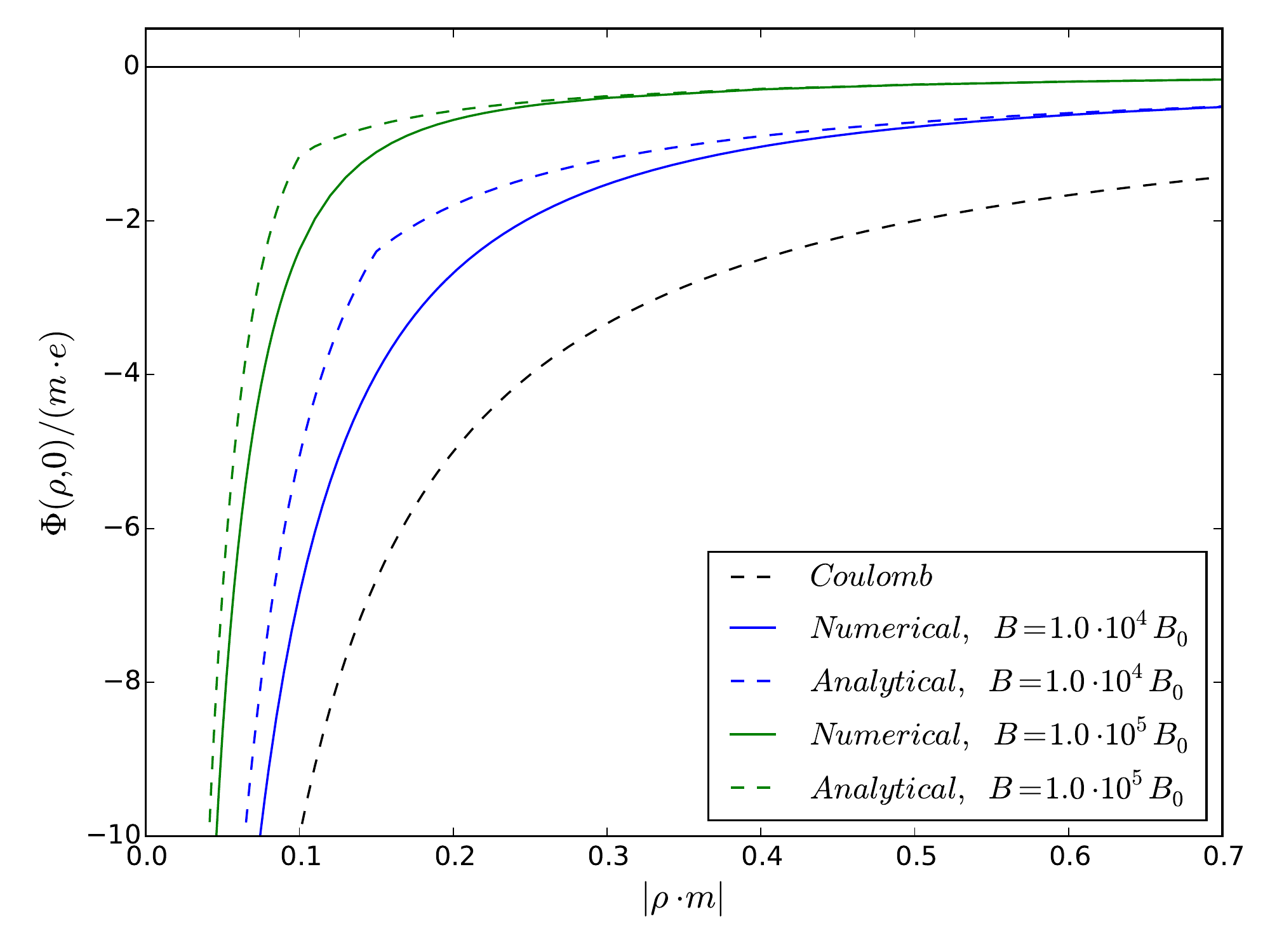}
    \caption{$\Phi\left(\rho,0\right)$}
    \label{fig:phi_rho}
  \end{subfigure}
  \caption{Numerical and analytical results for $B=10^{4}B_{0}$ and
    $B=10^{5}B_{0}$. The lowest dashed line corresponds to the Coulomb
    potential; two blue lines (solid and dashed) above the Coulomb
    potential correspond to numerical and analytical results for
    $B=10^{4}B_{0}$; two upper lines (solid and dashed green)
    correspond to numerical and analytical results for
    $B=10^{5}B_{0}$.}
  \label{fig:phi_z_rho}
\end{figure}

Evaluating this exact integral we can check the precision of
analytical estimations. With (\ref{eq:Phi_z}) we obtain the analytical
estimate for $\Delta\Phi\left(0,0\right)$:
\begin{equation}
  \label{eq:Phi_0_0}
  \Delta\Phi^{\rm analyt}\left(0,0\right)=\lim\limits_{z\to 0}\frac{e^{-z\sqrt{6m^{2}}}-e^{-z\sqrt{(2/\pi)e^{3}B+6m^{2}}}}{z}=\sqrt{(2/\pi)e^{3}B+6m^{2}}-\sqrt{6m^{2}}.
\end{equation}
For $B=10^{4}B_{0}$ we get
$\Delta\Phi^{\rm analyt}\left(0,0\right)/m\approx 4.793$. The
numerical value of exact integration is 4.41692858. We see that at
this point the analytical solution is quiet close to the numerical
one. This agreement is further confirmed in Fig. \ref{fig:phi_z},
where both numerical and analytical results are shown for
$\Phi\left(0,z\right)$. One can see a very good agreement between
numerical and analytical results for any $z$.

In Fig. \ref{fig:phi_rho} both numerical and analytical results are
shown for $\Phi\left(\rho,0\right)$. The agreement is not so good as
for $\Delta\Phi\left(0,z\right)$.

\begin{figure}[ht!]
  \centering
  \includegraphics[width=0.5\textwidth]{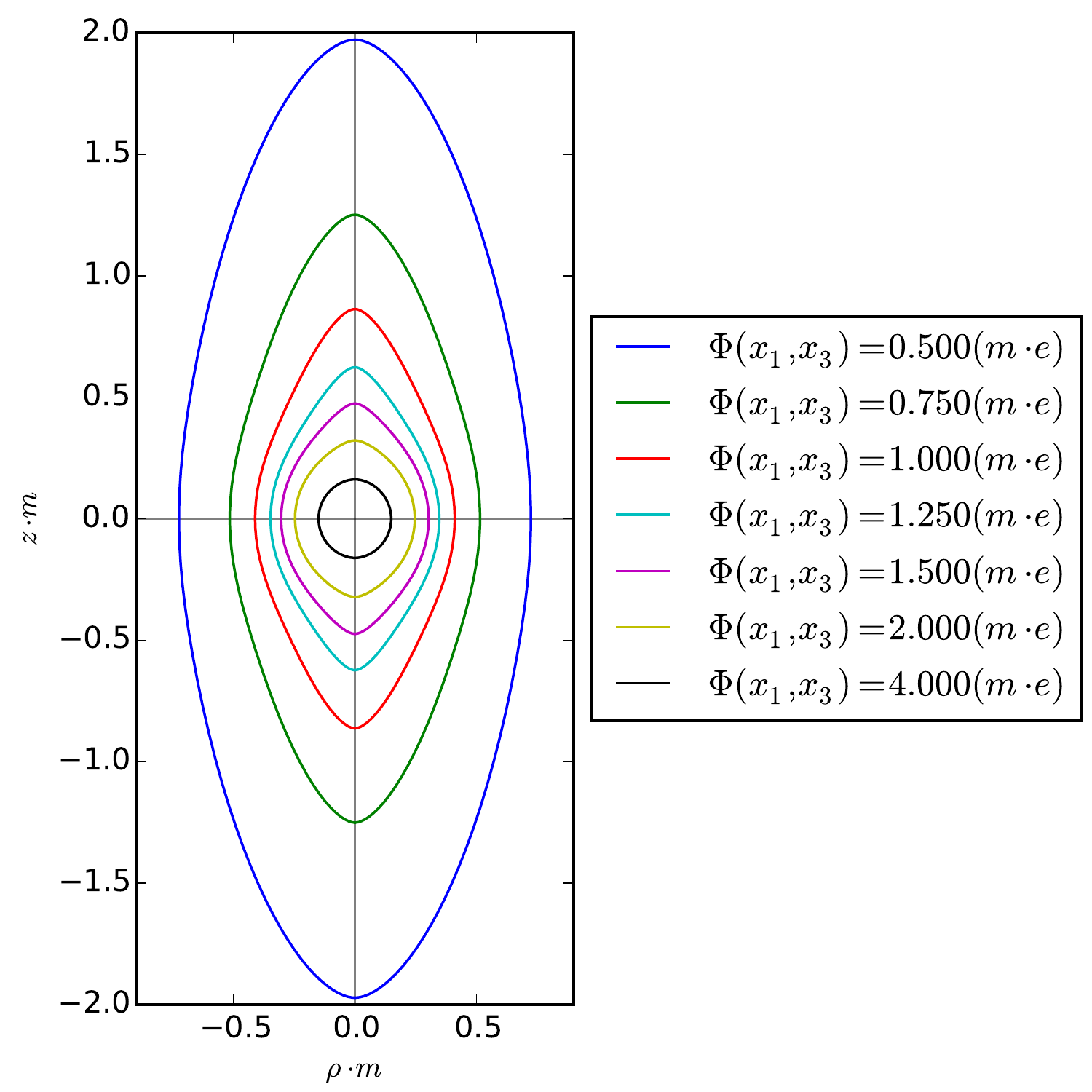}
  \caption{The equipotential lines for $B=10^{4}B_{0}$. The values in
    the legend correspond to the equipotential lines from outer to the
    inner one.}
  \label{fig:lines_1e4}
\end{figure}
\begin{figure}[ht!]
  \centering
  \begin{subfigure}[b]{0.45\textwidth}
    \includegraphics[width=\textwidth]{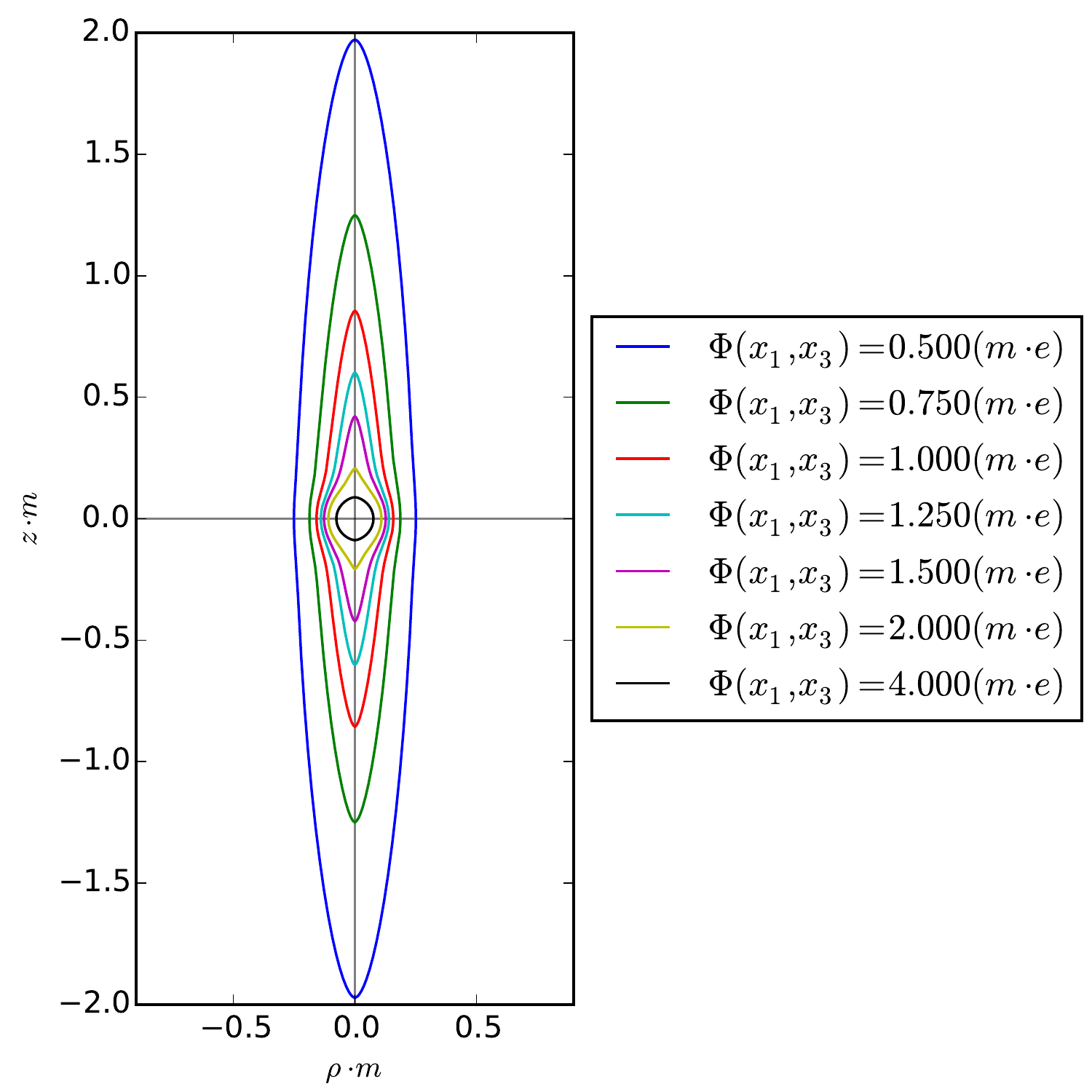}
    \caption{Overview}
    \label{fig:lines_1e5_overview}
  \end{subfigure}
  \begin{subfigure}[b]{0.45\textwidth}
    \includegraphics[width=\textwidth]{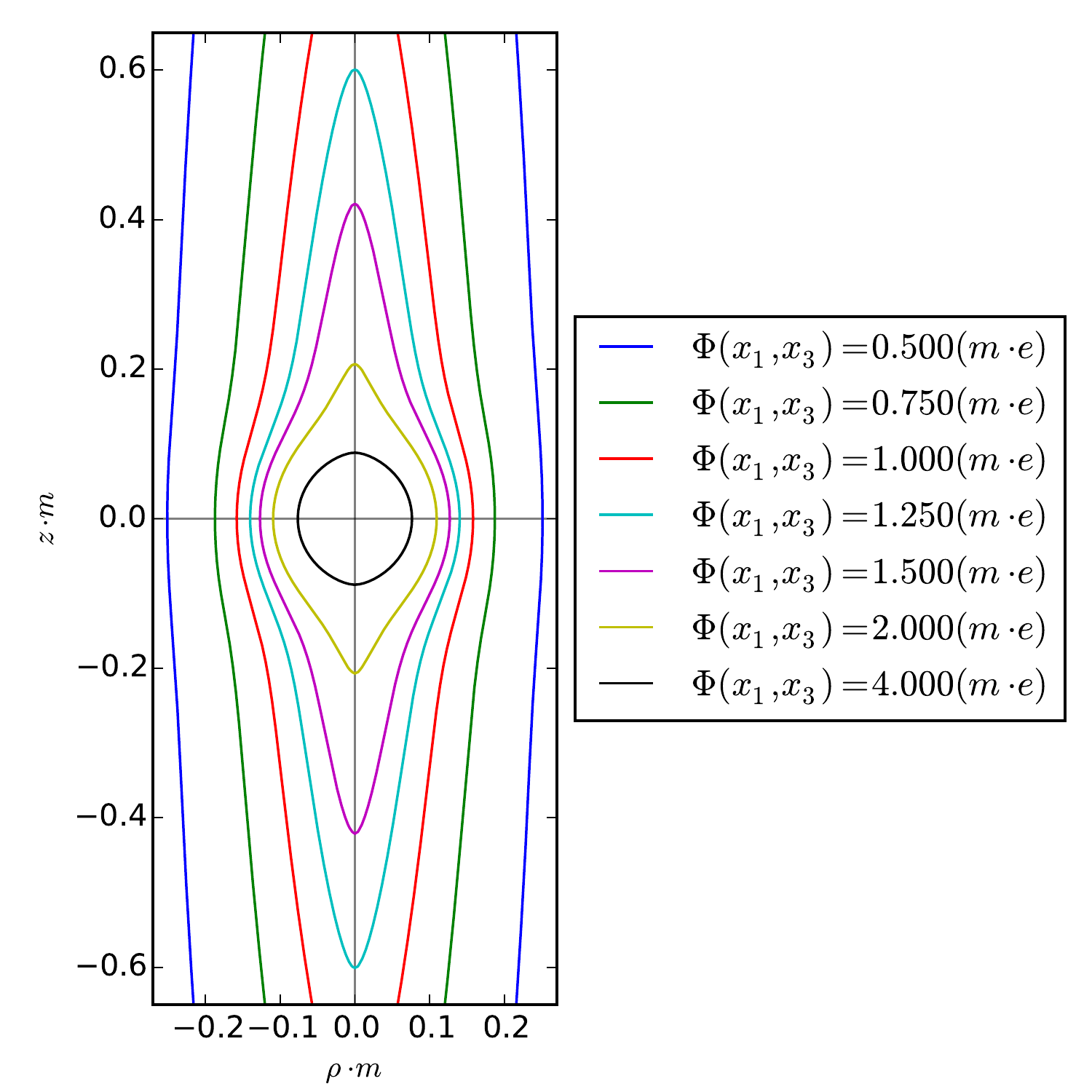}
    \caption{Central part}
    \label{fig:lines_1e5_center}
  \end{subfigure}
  \caption{The equipotential lines for $B=10^{5}B_{0}$. The legend for lines is the same as in Fig.~\ref{fig:lines_1e4}.}
  \label{fig:lines_1e5}
\end{figure}

Using numerical results we can obtain the correct spatial structure of
the potential. The equipotential lines for $B=10^{4}B_{0}$ and
$B=10^{5}B_{0}$ are shown in Fig.~\ref{fig:lines_1e4} and
Fig.~\ref{fig:lines_1e5} correspondingly. The central part of the
pattern shown in Fig.~\ref{fig:lines_1e5_overview} is magnified in
Fig.~\ref{fig:lines_1e5_center}. We see that outer equipotential line
has elliptic shape while inner lines are ``eye-shaped'' rather than
elliptic. It means that the modified potential of the pointlike
charge diminishes with $\rho$ faster than it was expected from
previous studies. If we consider the nucleus with finite radius
instead of the pointlike charge, we obtain that the potential
magnitude along $\rho=0$ line will be smaller for mid-range distances,
$z\lesssim 1/m$, than in the case of elliptic lines (since the
contribution to the potential from the charges dislocated from
$\rho=0$ will be smaller). It can be easily seen in the following
configuration: let us consider the charge distributed along the ring
$\rho=\rho_{0}=\mathrm{const}$ in the $z=0$ plane. In this case the
potential along $z$ axis ($\rho=0$) will be the same as the pointlike
charge potential along $\rho=\rho_{0}$ line. Since the pointlike
charge potential diminishes with $\rho$ faster than in the case of
elliptic lines, it proves the claim from above.

Let us consider an electron in the field of a nucleus with a finite
size in the presence of a superstrong magnetic field. The electron is
localized in the direction transverse to the magnetic field at
distances $\sim a_{H}\equiv 1/\sqrt{eB}$. When the magnetic field is
so strong that $a_{H}$ gets smaller than the nucleus size, the charge
distribution becomes important. As we have shown above the potential
is weaker than it was expected. It means that energy levels in the
field of such a nucleus will be higher. The detailed investigation of
this problem is a subject for a separate study.
 
\section{Conclusions}
\label{sec:conclusions}

The paper considers the Coulomb potential in a superstrong external
magnetic field. Due to the enhancement of loop corrections the
potential is modified (screened). The results for these corrections
were known from previous studies, but not in the whole space. In this
paper we calculated numerically the modified potential in all
space. With the help of numerical results we estimated the precision
of analytical results and obtained some new features. It turned out
that the equipotential lines are not ellipses in the mid-range
distances, $z\lesssim 1/m$, see Fig. \ref{fig:lines_1e4},
\ref{fig:lines_1e5}. It means that potential diminishes with $\rho$
faster than it was expected from previous studies. This feature may be
important for some problems, e.g. with spatially distributed charges
(like energy levels calculation for atoms and ions with high-Z
nuclei).

Authors are grateful to M. Vysotsky for valuable remarks and
discussions.

Authors are supported by RFBR under grant 16-32-00241, by the Grant of
President of Russian Federation for the leading scientific Schools of
Russian Federation, NSh-9022-2016, and by ``Dynasty
Foundation''. Sergey Godunov is also supported by RFBR under grants
16-32-60115 and 16-02-00342.


\begin{thebibliography}{99}
\bibitem{ShU-1}
A.E. Shabad, V.V. Usov, Phys. Rev. Lett. {\bf 98} (2007) 180403.
\bibitem{ShU-2}
A.E. Shabad, V.V. Usov, Phys. Rev. D {\bf 77} (2008) 025001.
\bibitem{Vysotsky-2010}
M.I. Vysotsky, Pis'ma Zh. Eksp. Teor. Fiz. {\bf 92} (2010) 22
[JETP Lett. {\bf 92} (2010) 15].
\bibitem{MV}
B. Machet, M.I. Vysotsky, Phys. Rev. D {\bf 83} (2011) 025022.
\bibitem{GMV}
S.I. Godunov, B. Machet, M.I. Vysotsky, Phys. Rev. D {\bf 85} (2012)
044058.
\bibitem{LS} Yu.M. Loskutov, V.V. Skobelev, Phys. Lett. A \textbf{56} (1976)
  151.
\bibitem{Godunov-2loop} S.I. Godunov, Phys. Atom. Nucl. \textbf{76}
  (2013) 901.
\bibitem{GSL} M. Galassi et al, GNU Scientific Library Reference
  Manual (3rd Ed.), ISBN 0954612078.
\bibitem{FFTW} The Fastest Fourier Transform in the West, http://www.fftw.org/
\end{thebibliography}
\end{document}